# High efficient metasurface quarter-wave plate with wavefront engineering

*Chen Chen[1,2], Shenglun Gao[1,2], Xingjian Xiao[1,2], Xin Ye[1,2], Shengjie Wu[1,2], Wange Song[1,2], Hanmeng Li[1,2], Shining Zhu[1,2], and Tao Li[1,2,*]*

[1] National Laboratory of Solid State Microstructures, Key Laboratory of Intelligent Optical Sensing and Manipulation, Jiangsu Key Laboratory of Artificial Functional Materials, College of Engineering and Applied Sciences, Nanjing University, Nanjing, China, 210093
[2] Collaborative Innovation Center of Advanced Microstructures, Nanjing, China, 210093

*Corresponding authors:*
*Prof. Tao Li, Email: taoli@nju.edu.cn Tel: +86-25-83593805 URL: http://dsl.nju.edu.cn/litao/*

**Abstract**: Metasurfaces with local phase tuning by subwavelength elements promise unprecedented possibilities for ultra-thin and multifunctional optical devices, in which geometric phase design is widely used due to its resonant-free and large tolerance in fabrications. By arranging the orientations of anisotropic nano-antennas, the geometric phase-based metasurfaces can convert the incident spin light to its orthogonal state, and enable flexible wavefront engineering together with the function of a half-wave plate. Here, by incorporating the propagation phase, we realize another important optical device of quarter-wave plate together with the wavefront engineering as well, which is implemented by controlling both the cross- and co-polarized light simultaneously with a singlet metasurface. Highly efficient conversion of the spin light to a variety of linearly polarized light are obtained for meta-holograms, metalens focusing and imaging in blue light region. Our work provides a new strategy for efficient metasurfaces with both phase and polarization control, and enriches the functionalities of metasurface devices for wider application scenarios.

## 1. Introduction

The emergence of metasurfaces with ultrathin, ultralight, and multifunctional advantages has spurred significant interest in light manipulations.[1-7] By judiciously designing the field discontinuities across the interface with different meta-atoms, full control of sophisticated wavefronts, polarization states, and their energy allocations can be achieved.[8-10] Plentiful functionalities with great potential in applications are demonstrated, such as metalens [11-15], meta-holograms,[16-18] and polarizers,[19-21] to name a few. Among them, simultaneous control of the polarization and phase plays a vital role and has already aroused numerous researches to explore its full potential.[22-26] Attempts such as utilizing two plasmonic nanopillars per period with different distance and orientation angle only work for oblique incidence.[27] Another method is exploiting the superposition of the two output circular polarization (CP) beams through two sets of nanopillars with different dimensions and starting orientation angles under linear polarization (LP) incidence.[28] Both of them are based on super unit cells with spatial superposition of inclusions, which would result in lower efficiency, inferior image quality, and lower space-bandwidth product. Recently, it is of great interest to combine the propagation phase and geometric phase (i.e., Pancharatnam–Berry (PB) phase) to realize full control of the polarization and phase on a single subwavelength unit cell.[29-31] However, they are usually focused on the polarization multiplexing to enhance the information capability, e.g., different functionalities are encoded with different polarization states. In fact, incorporating the same wavefront engineering with different polarization states to realize specific polarization conversion is quite useful but remains rarely explored. As a basic optical component, quarter waveplate (QWP) (normally convert the CP light to LP light and vice versa) plays an important role in light manipulation.[32,33] It would be highly desirable to find ways to implement QWPs on a single metasurface.

Here, we provide a straightforward design principle for metasurfaces (e.g. meta-holograms and metalens) to achieve the QWP functionality by utilizing both the co- and cross-polarized spin light. By combing the propagation phase with PB phase, we first demonstrate the modulate capacity with spin-selected holographic images (the commonly unmodulated co-polarized light and cross-polarized light) with SiNx metasurfaces in the visible spectrum. We further realize the QWPs with wave manipulation abilities by controlling the superposition of two output CP light. Other elliptical polarizations with designed wavefront are also produced experimentally.

The polarization reconstruction and wave manipulation based on two orthogonal CP bases certainly expands the practical application possibilities and could trigger versatile function integrations for advanced compact systems.

**2. The Design Principles**

PB phase-based metasurfaces can achieve a full phase control by adjusting the orientation angle of the meta-atoms with identical geometry.[34] The cross-polarized light will have extra $\mp i2\sigma\theta$ phase modulation under normal CP light incidence, where $\theta$ is the rotation angle from the $x$-axis, and $\sigma$ indicates the handness of the CP light. However, the co-polarized scattered light is usually ignored, which unavoidably results in background noises or a dazzling spot in the hologram image.[28,31,35,36] In order to surmount this restriction, we propose to modulate the co- and cross-polarized light independently by combining the propagation phase and PB phase. The total Jones matrix describing the relation between the input electric field ($E_{in}$) and the output electric field ($E_{out}$) in a circular base can be written as:

$$J=R_c(-\theta)\begin{bmatrix}e^{i\phi_{RR}} & e^{i\phi_{RL}} \\ e^{i\phi_{LR}} & e^{i\phi_{LL}}\end{bmatrix}R_c(\theta)=\begin{bmatrix}e^{i\phi_{RR}} & e^{i(2\theta+\phi_{RL})} \\ e^{i(-2\theta+\phi_{LR})} & e^{i\phi_{LL}}\end{bmatrix}, \quad (1)$$

where $R_c(\theta)$ is the rotation matrix, $\phi_{RL}$, $\phi_{LR}$, $\phi_{RR}$ and $\phi_{LL}$ is the propagation phase. In this work, we consider the widely used rectangular nanopillars, so the phase shift $\phi_{RR}=\phi_{LL}$ and $\phi_{RL}=\phi_{LR}$ due to the mirror symmetry. In this case, if the incident wave is right circularly-polarized (RCP), the output electric field $E_{out}$ becomes:

$$E_{out}=JE_{in}=\begin{bmatrix}e^{i\phi_{RR}} & e^{i(2\theta+\phi_{RL})} \\ e^{i(-2\theta+\phi_{RL})} & e^{i\phi_{RR}}\end{bmatrix}\begin{bmatrix}1 \\ 0\end{bmatrix}=\begin{bmatrix}e^{i\phi_{RR}} \\ e^{i(-2\theta+\phi_{RL})}\end{bmatrix}. \quad (2)$$

From Eq. (2), we can clearly conduct the phase modulation for the transmitted/reflected RCP and LCP wave independently. For example, if the transmitted/reflected RCP light is designed for a special function with the phase profile $\varphi_1(x,y)$ and the LCP light with another phase profile $\varphi_2(x,y)$, then $\varphi_1(x,y)=\phi_{RR}(x,y)$, and $\varphi_2(x,y)=-2\theta(x,y)+\phi_{RL}(x,y)$. It means the phase manipulation can be implemented by arranging different nanopillars with different rotation angles in different positions. Moreover, if a reference phase ($\varphi=-2\theta_0$) is added to $\varphi_2(x,y)$, representing all the nanopillars rotating $\theta_0$ extra angle, which does not influence the wave manipulation capacity, but it surely affects the output polarization state considering the superposition of the other CP light. The schematics of the electric field changing process (RCP beam |R> passing through a rotated nanopillar) is shown in **Figure 1(a)**, the output two beams can be written as $a_R e^{i\phi_{RR}}$|R> and $a_L e^{i(\phi_{RL}-2\theta-2\theta_0)}$|L>, where $a_R$ and $a_L$ are the corresponding amplitudes of the output RCP and LCP beams. Ignoring the material loss, the amplitude distributions of the two output CP light are shown in **Figure 1(c)**, relevant to the specific structure parameters and complementary to each other. The above mentioned reference phase $\varphi$ is related to the total extra rotation angle $\theta_0$, with its relationship shown in **Figure 1(d)**. If the wave manipulations of the two CP light are the same with each other ($\phi_{RR}=\phi_{RL}-2\theta$), then the superimposed polarization state can be written as |n>=$a_R$|R>+ $a_L e^{i\varphi}$|L>, where $a_R/a_L$ and $\varphi$ can be independently modulated. In this way, as shown in **Figure 1(b)**, an arbitrary polarization state can be reconstructed by changing $a_R/a_L$ (shifting along the longitude direction) and $\varphi$ (moving along latitude direction), covering the whole Poincaré sphere. The azimuth angle $\psi$ and ellipticity angle $\chi$ can be derived as $\psi=\theta_0$ and $\chi=\frac{1}{2}\arcsin\frac{a_R^2-a_L^2}{a_R^2+a_L^2}$.

## 3. Experimental Results

### 3.1. Phase modulation capability

As a proof of concept, we consider the Silicon nitride (SiNx) metasurfaces consisting of nanopillars with different shapes covered on the fused-silica substrate. SiNx was chosen because of its low loss in visible light and compatibility with CMOS processes. The design wavelength is 470 nm and the metasurface works in a transmitted way. As illustrated in **Figure 1(a)**, the unit cell period is chosen as 300 nm satisfying the Nyquist theory.[37] The nanopillar height is set as 800 nm to provide phase change covering 0~2π. **Figure 2(a)** shows the simulated efficiency distribution of nanopillars with different widths and lengths for the transmitted RCP and LCP light with RCP light incidence. The simulated phase responses of the nanopillars for the working RCP ($\phi_{RR}$) and LCP ($\phi_{RL}$) light are shown in **Figure 2(b)** and **2(c)**, respectively. The purple dots mark the selected antennas with required phase responses and homogenous amplitudes (to satisfy $a_R/a_L=1$). **Figure 2(d)** illustrates the specific amplitude distribution, the average amplitude for RCP light is $a_R$=0.47, and for LCP light is $a_L$=0.46, nearly equal to each other. It is worth mentioning that there is still a large parameter space for the phase modulation while restricting the amplitude distribution, especially compared to the high $a_L$ section when purely utilizing the PB phase (see **Supplementary Figure S1**).

To verify the phase modulation ability, we first demonstrated an independent spin polarization hologram metasurface with RCP light incidence. The transmitted RCP light is modulated to produce a far-field hologram image with "NJU" based on propagation phase ($\phi_{RR}$) and LCP light is manipulated to present a representative building of Nanjing University (i.e. a 600-years building named Bei-Da) in the far field based on propagation phase ($\phi_{RR}$) and PB phase ($\phi_{RL}$-2$\theta$). The optimized phase profiles are based on Gerchberg-Saxton algorithm.[38-40] The metasurface with a footprint of 150 μm × 150 μm is fabricated using a conventional nanofabrication process (see **Experimental Section**) and its scanning electron microscopy (SEM) image is shown in **Figure 2(e)**. The experimental light intensity profiles for the transmitted RCP and LCP light in the projected plane are shown in **Figure 2(f)**, which are consistent with the designed images and have negligible zero spots.

### 3.2. Meta-holograms with QWP effect

After verifying the phase modulation capability, we further demonstrate the polarization state manipulation. The output RCP and LCP light are designed with the same function (e.g. the same focal length or the same hologram image) by modulating the propagation phase $\phi_{RR}(x,y)$ and the compensatory PB phase -2$\theta(x,y)$. To realize the QWP functionality, we choose the same nanopillars as marked in **Figure 2(b)** to obtain equal $a_R$ and $a_L$. The reference phase $\varphi$ (related to the extra rotation angle $\theta_0$) is modulated to get different LP light. As shown in **Figure 3(a)**, the red dots on the Poincaré sphere mark the designed output linear polarization state under RCP incidence. The SEM image of the metasurface with a footprint of 150 μm × 150 μm for $x$-polarized (dot A) hologram is shown in **Figure 3(b)**. The insert picture is the enlarged image. The directly (without analyzer) measured holographic image is shown in **Figure 3(c)**. Due to the utilization of both the co- and cross-polarized light, the middle zero spot is nearly negligible, indicating a very high diffraction efficiency (98%). The zero spot is difficult to completely disappear because of k-space imaging of the light passing through the gap between nanopillars. To verify its polarization properties, we add a polarizer to detect the relative intensity profile. When the polarizer is set as 0° (the transmission axis is parallel to the $x$-axis), the hologram intensity profile (see **Figure 3(d)**) is nearly the same as the measured image without analyzer (**Figure 3(c)**). When the polarizer is rotated as 45° and 90°, the intensity of the output hologram image change from attenuation to disappearance that surely verifies the linear polarization ($x$-polarized) properties (see **Figure 3(e)** and **3(f)**), indicating the functionality of QWP together with wave manipulation. With different designed reference phase $\varphi$ (the total extra rotation angle $\theta_0$), other output LP states can also be produced. **Figure 3(g)-(i)** show the measurements (without analyzer) of the other meta-QWPs (RCP to B:45°, C:90° and D:-45° LP) for holographic imaging. The respective insert SEM image, compared to the insert one in **Figure 3(b)**, reveals the different extra rotation angle of the whole nanopillars. Thus, hologram images with different LP states under RCP incidence can be obtained by rotating the metasurface nanopillars.

### 3.3. Metalens with QWP effect

To further demonstrate the manipulation capability, we design a meta-QWP for focusing and imaging. As illustrated in **Figure 4(a)**, the designed single-layer metasurface can act as a QWP and a lens at the same time. The phase profile follows[41] $\varphi_1(x,y)=\varphi_2(x,y)=\frac{2\pi}{\lambda}\left(f-\sqrt{x^2+y^2+f^2}\right)$, where the focal length $f$=100 μm. **Figure 4(b)** shows the SEM image of the fabricated metalens (D=150 μm, NA=0.6) with $x$-polarization output under RCP incidence (the detailed analysis are provided in **Supplementary Figure S2**). As demonstrated above, other LP output can also be obtained as long as rotating the metasurface. The directly measured focus spot is shown in **Figure 4(c)**, with the full width at half maximum (FWHM) of 477 nm. The focusing efficiency (the ratio of the optical power in an area of diameter 3×FWHM to the total transmitted power) is calculated as 53%. When incidence is switched to LCP light, the focus becomes $y$-polarized due to the superposition of the output LCP and RCP components, as shown in **Figure 4(d)** with FWHM of 443 nm. Similarly, when the illumination is $x$-polarized, the focus (FWHM is measured as 630 nm) is accordingly LCP (see **Figure 4(e)**), and the focus changes to RCP when the incidence is $y$-polarized (see **Figure 4(f)**) with the FWHM of 630nm as well. If the front and back of the metasurface is flipped, the polarization changing rule will also be reversed accordingly. As shown in **Figure 4(g)**, when the $x$-polarized light is illuminated from the other side, the focus becomes RCP with the measured FWHM of 614 nm. The difference of FWHM in these situations is mainly due to the measurement error in the experiment. Here, simulations of the metalens with the same *NA* (0.6) are also performed to verify the focusing properties (**Supplementary Figure S3**). The simulated FWHM is 400 nm and the focusing efficiency is 77%. The experimental quantitative results are not as good as the simulation ones due to the imperfections of the fabricated nanopillars. Yet qualitative performances such as the polarization state properties are in high accordance with the simulations. Furthermore, we conduct the imaging test of the fabricated sample. The polarization of the image is consistent with the polarization of the focus. **Figure S4** shows the image in the situation of **Figure 4(c)** (RCP incidence) without analyzer and with 90° polarizer, and the extinction properties indicate the $x$-polarization the image is. Note that the metasurface can also produce a good image regardless of the additional optical elements for generating and detecting the corresponding polarization, as shown in **Figure 4(h)** and **4(i)**. The right side of the **Figure 4(h)** is the enlarged central part image of the 1951 United States Air Force (USAF) resolution test chart, manifesting clear resolution of Element 1 and Group 9 (i.e., a resolution of 980 nm). **Figure 4(i)** is the image of chicken erythrocytes, and the nuclei are clearly distinguishable.

### 3.3. Metalens with elliptic polarizations generation

In addition, this method can be scaled to other polarization states, such as elliptic polarizations E and F on the Poincaré sphere (see **Figure 5(a)**). Specifically, for the polarization state E, $a_R/a_L=1/2$ and $\varphi=0$, the amplitude distribution of the selected nanopillars are shown in **Figure 5(b)**, with the average $a_R=0.30$ and $a_L=0.61$ ($a_R+a_L<1$ is due to the intrinsic material loss), and the designed focal length is set as 200 μm with $NA=0.35$. **Figure 5(c)** shows the intensity profile of the focus with the FWHM of 1.18 μm and the focusing efficiency of 81% (polarization analysis results can be found in **Figure S5**). **Figure 5(d)** displays the image (without analyzers) of the ascaris eggs directly illuminated by a halogen lamp with a filter of 470 nm, clearly showing the elliptical shapes. Similarly, for the polarization state F, $a_R/a_L=7/2$ and $\varphi=\pi$, and the corresponding amplitude distribution of the selected antennas are also shown in **Figure 5(e)**, with the average amplitude $a_R=0.71$ and $a_L=0.23$. The focal length is designed as 300 μm with $NA=0.24$. Enlarged focal spot with the FWHM of 1.60 μm is shown in **Figure 5(f)** and the focusing efficiency is calculated as 87%. **Figure 5(g)** illustrates the image of Sector Star Target which resolution at the central circle is 8.7 μm. SEM images of these samples can be found in **Figure S6.**

### 4. Conclusion

In conclusion, we have demonstrated a straightforward method for realizing phase manipulation and polarization control at the same time. Hologram images with negligible zero spots are obtained with high diffraction efficiencies (98%), and the linear polarization properties are also verified. Metalens as a QWP is demonstrated experimentally as well. This method can be extended to acting as other wave-plates, for example, RCP to elliptic polarizations are also validated likewise. The utilization of the normally ignored co-polarized spin light surely enhances the device performances and increases the functionalities. With the proper choice of materials and scaling of the designs, the method can be generalized to other wavelengths. This research extends and strengthens the practical applications of metasurfaces with phase and polarization manipulation.

## 5. Experimental Section

*Numerical simulations:* The simulated material parameters of SiNx is adopted from the experimental measurement. The wavelength is fixed at 470 nm, and the refractive index is $n=2.032094+0.000309i$. Full-wave simulations are carried out using commercial finite-difference time domain (FDTD) software, Lumerical. For individual nanopillars, perfectly matched layer (PML) conditions in the direction of the light propagation and periodic boundary conditions along all the in-plane directions were used. For the 3D metalens simulation, PML conditions were used in all the directions.

*Fabrication of the metasurfaces:* The silicon nitride layer was deposited on the fused-silica substrate using the plasma enhanced chemical vapor deposition (PECVD) to a final thickness of 800 nm. Then 200 nm PMMA A4 resist film was spin-coated onto the substrate and baked at 170°C for 5 min. Next, a 42 nm thick layer of a water-soluble conductive polymer (AR-PC 5090) was spin-coated on the resist for the dissipation of E-beam charges. The device pattern was written on an electron beam resist using E-beam writer (Elionix, ELS-F125). The conductive polymer was then dissolved in water and the resist was developed in a resist developer solution. An electron beam evaporated chromium layer was used to reverse the generated pattern with a lift-off process, and was then used as a hard mask for dry etching the silicon nitride layer. The dry etching was performed in a mixture of CHF3 and SF6 plasmas using an inductively coupled plasma reactive ion etching process (Oxford Instruments, PlasmaPro100 Cobra300). Finally, the chromium layer was removed by the stripping solution (Ceric ammonium nitrate).

*Optical measurement:* A white-light laser (Fianium Super-continuum, 4W) with a 470 nm filter which bandwidth is 10 nm was used as the illumination source for holographic imaging and focus measurement. For the lens imaging, a halogen lamp was used for the illumination. A polarizer (Thorlabs, WP25L-VIS) and a QWP (Thorlabs, AQWP05M-600) were used for generating required polarization incidence and corresponding analyses, and an objective (NA=0.9, 100X) was used to collect the image.

**Acknowledgements**
We acknowledge the micro-fabrication center of the National Laboratory of Solid State Microstructures (NLSSM) for technique support. This work was supported by the National Key R&D Program of China (2017YFA0303701, 2016YFA0202103), National Natural Science Foundation of China (Nos. 91850204, 11674167, 11621091), and PAPD from Jiangsu province. Tao Li thanks the support from Dengfeng Project B of Nanjing University.

**Figures:**

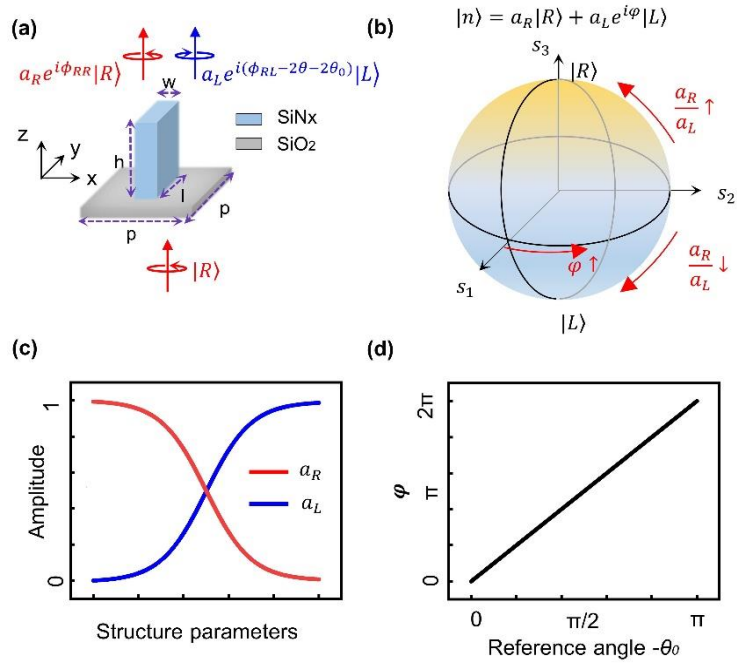

**Figure 1.** Schematics of the design principle. (a) Electric field changing process through a dielectric rotated nanopillar. (b) Poincaré sphere containing the full polarization states with different reference phase $\varphi$ and amplitudes $a_R$ and $a_L$. (c) Amplitude distribution with structure parameters. (d) Relationship of the reference phase $\varphi$ added to LCP and the total extra rotation angle $\theta_0$.

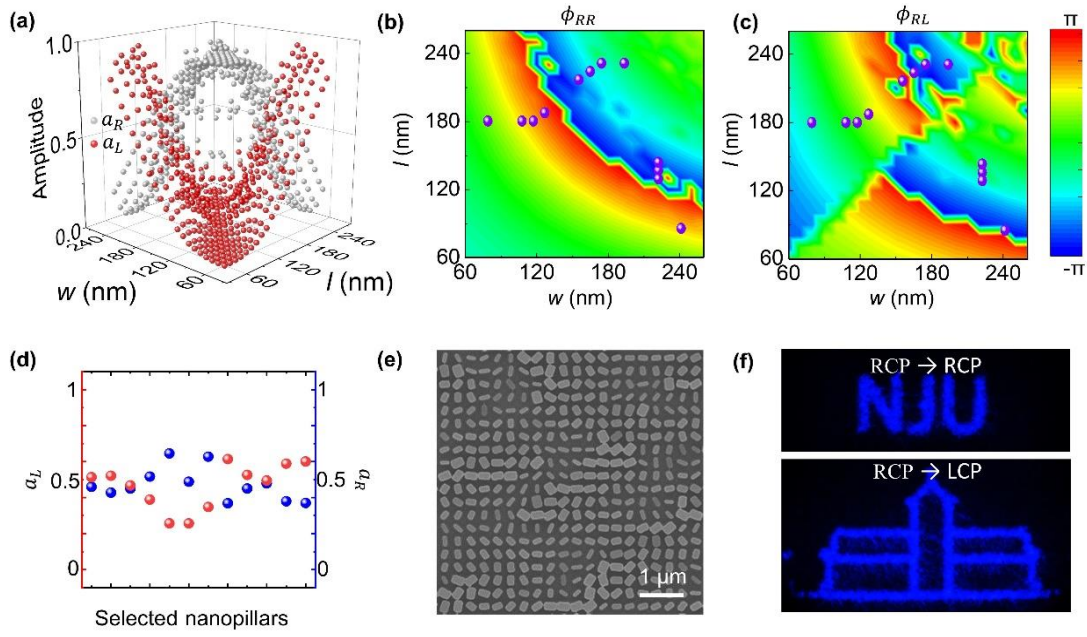

**Figure 2.** Independent wave manipulation of the two output CP beams. (a) Simulated amplitude distribution of the transmitted RCP and LCP light with the nanopillar's geometric parameters under RCP incidence. Phase responses of the transmitted RCP (b) and LCP (c) light with different nanopillars under RCP incidence. Purple dots mark the chosen antennas. (d) Amplitude distribution of the selected nanopillars. (e) SEM image of the fabricated metasurface. (f) Meta-hologram images with RCP and LCP analyzer.

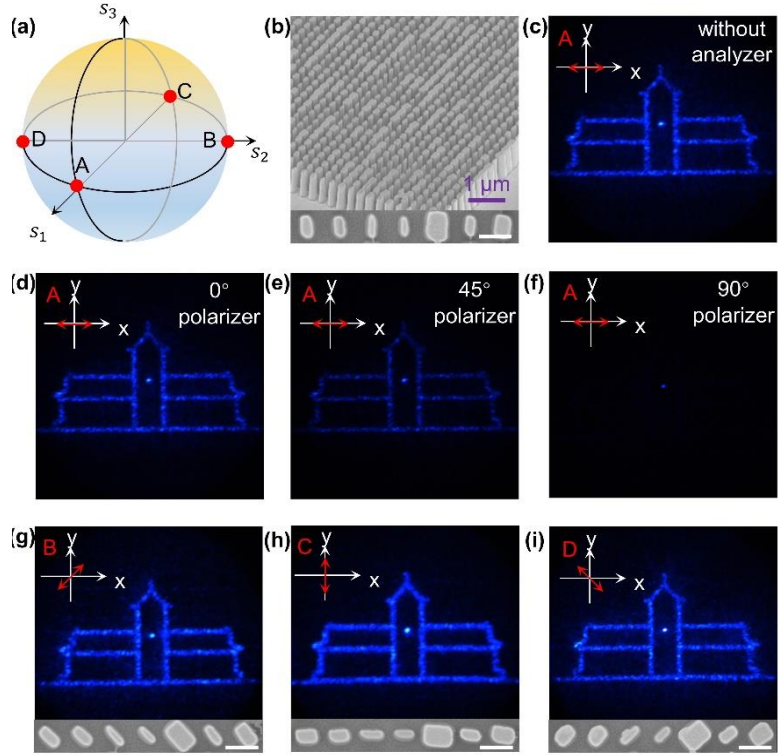

**Figure 3.** Different meta-QWPs for holographic imaging. (a) Red dots on the Poincaré sphere indicate the designed output LP with RCP incidence. (b) SEM images of the metasurface for *x*-polarized (A) output. Scale bar = 300 nm. (c) Meta hologram image (A: *x*-polarized) without analyzer. The intensity profiles of the hologram with a (d) 0° polarizer, (e) 45° polarizer and (f) 90° polarizer. Meta hologram images without analyzer for (g) B:45°-polarization, (h) B:90°-polarization and (i) B:-45°-polarization. Inserts are the corresponding SEM images. Scale bar = 300 nm.

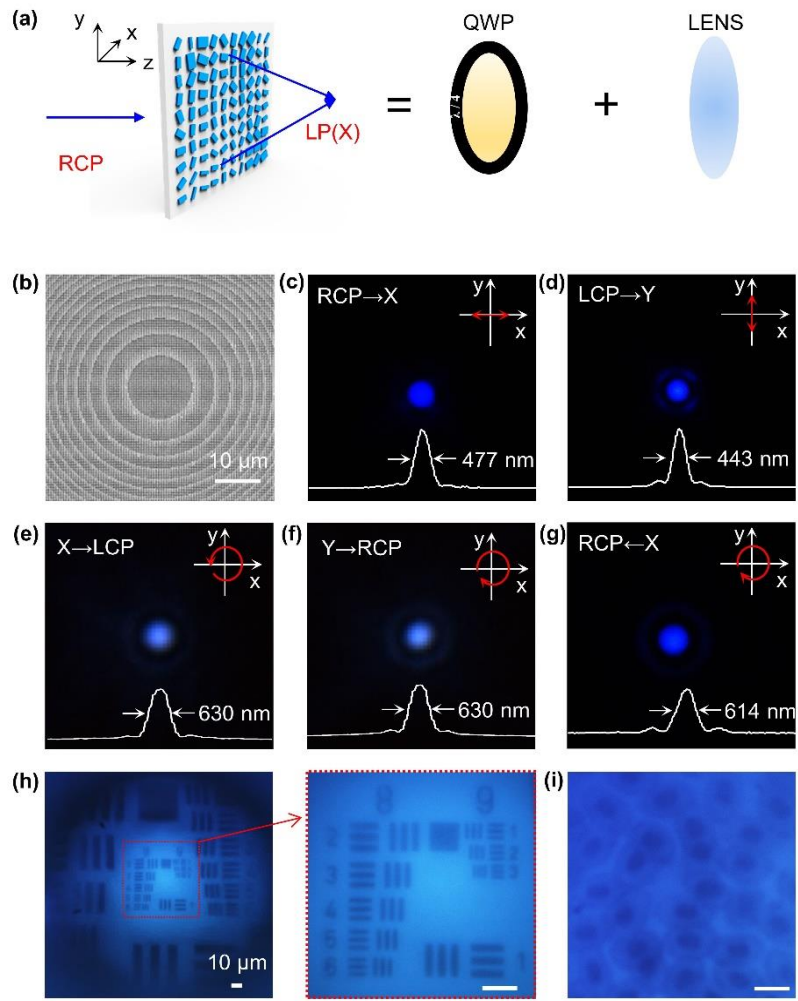

**Figure 4.** Meta-QWP for focusing and imaging. (a) Schematic illustration of the metasurface acting as the QWP and the lens at the same time. (b) SEM image of the fabricated metasurface. Intensity profiles of the focus when the incidence (from the substrate) is (c) RCP, (d) LCP, (e) *x*-polarized and (f) *y*-polarized. The intensity profile of the focus when *x*-polarized light is incident from the side of the nanopillars. Image of (h) the USAF resolution test chart and (i) chicken erythrocytes (without analyzer) when directly illuminated by a halogen lamp with a filter of 470 nm. Scale bar = 10 μm.

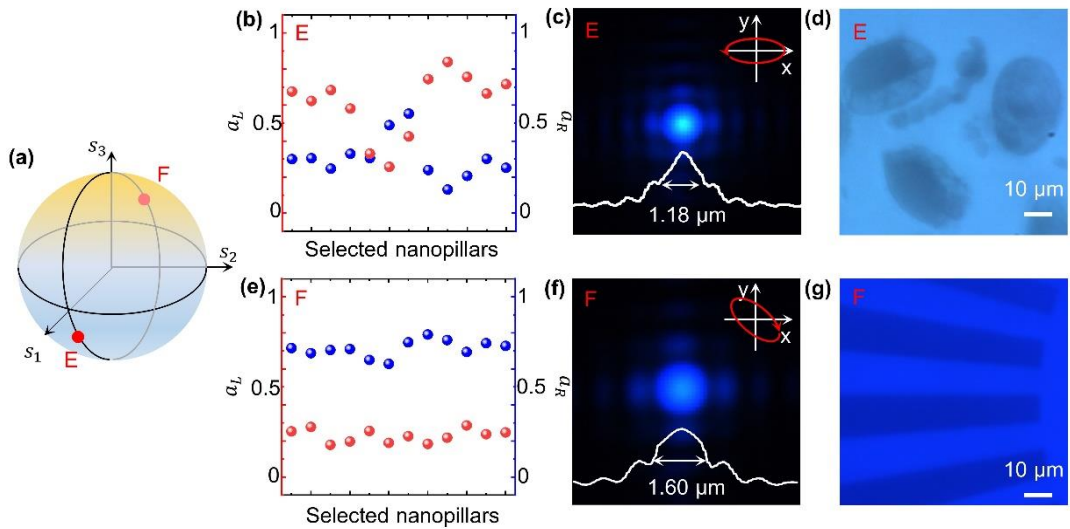

**Figure 5.** Realization of the meta-lens with elliptic polarization states. (a) Red dots mark the designed elliptic polarization states on the Poincaré sphere. Amplitude distribution for the elliptic polarization state (b) E and (e) F. Intensity profile of the focus with the elliptic polarizations state (c) E (NA=0.35) and (f) F (NA=0.24). (d) Direct image of the ascaris eggs with the metalens which has the elliptic polarization E focus. (g) Direct image of the Sector Star Target with the metalens which has the elliptic polarization F focus.

# Supporting Information

**High efficient metasurface quarter-wave plate with wavefront engineering**

*Chen Chen[1,2], Shenglun Gao[1,2], Xingjian Xiao[1,2], Xin Ye[1,2], Shengjie Wu[1,2], Wange Song[1,2], Hanmeng Li[1,2], Shining Zhu[1,2], and Tao Li[1,2],\**

**S1. Amplitude distribution and phase distribution.**

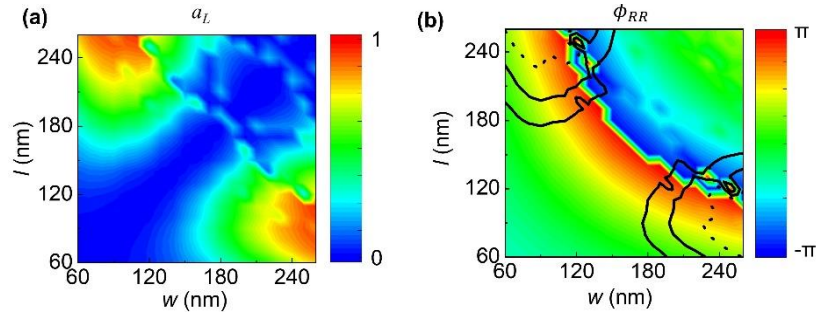

**Figure S1.** (a) Simulated amplitude distribution of $a_L$ under RCP incidence. (b) Phase distribution of the transmitted RCP light. The section enclosed by the dashed line in the corner marks the meta-atoms with $a_L \geq 80\%$, and the section enclosed by the solid line marks the meta-atoms with $40\% \leq a_L \leq 60\%$.

**S2. Polarization analyses of the linear polarized focus.**

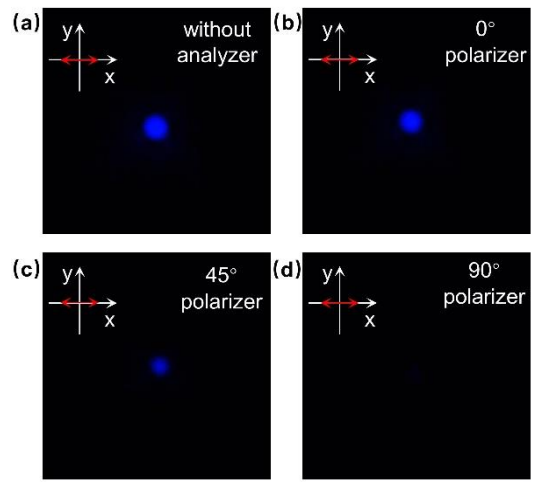

**Figure S2.** (a) Measured focal spot with linear polarization in x-axis direction. Measured intensity profiles of the focus with (b) a 0° polarizer, (c) a 45° polarizer, (d) a 90° polarizer.

## S3. Simulation results of the focal spot with different incidences.

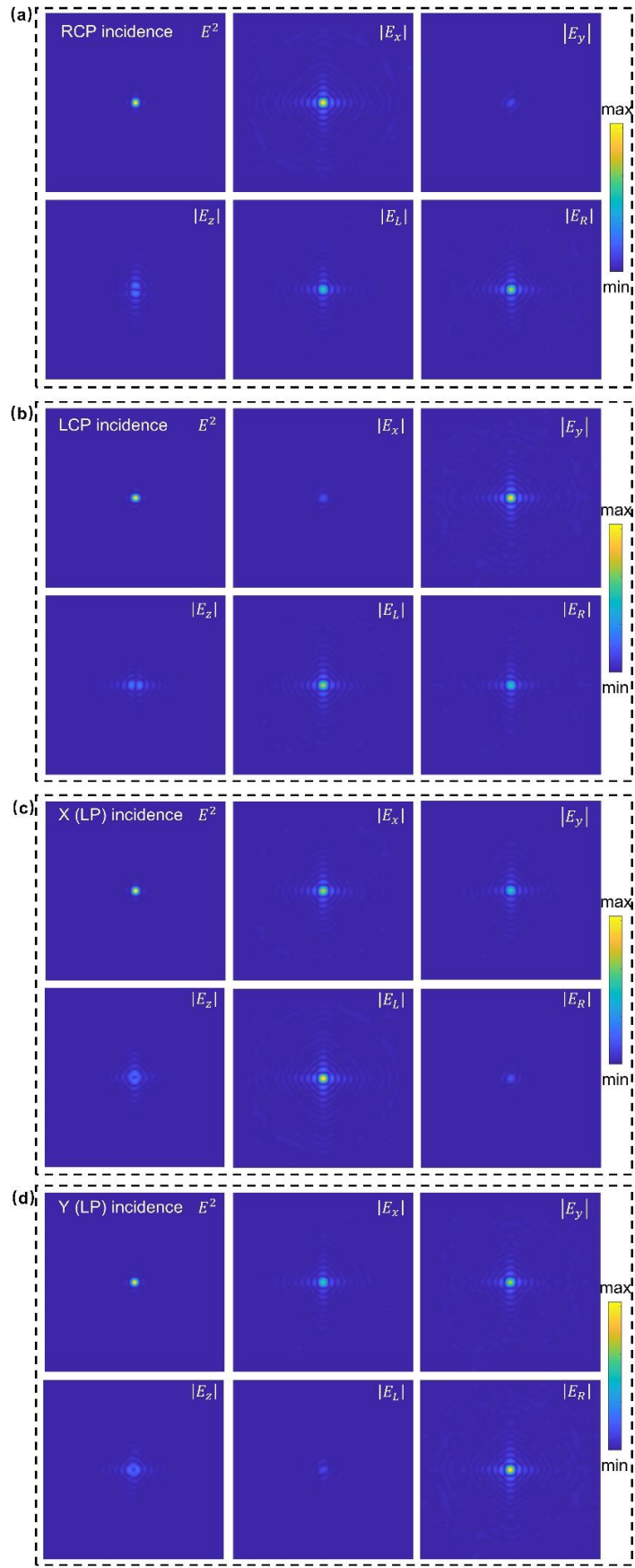

**Figure S3.** Different electric compontens of the focal spot when the metalens is illuminated by (a) RCP (b) LCP (c) *x*-polarized and (d) *y*-polarized light from the substrate.

### S4. Polarization analyses of image with the metalens.

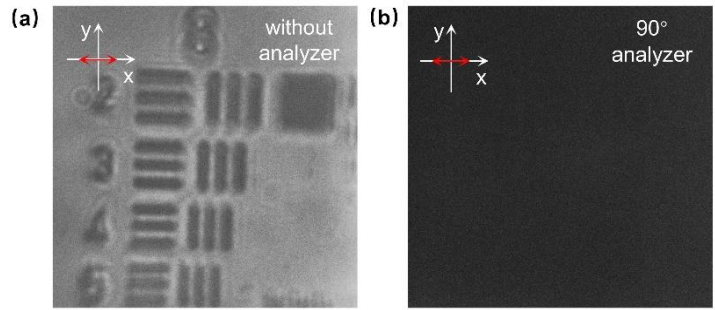

**Figure S4.** Image with the metalens corresponding to Figure 4(c), (a) without analyzer and (b) with 90° polarizer.

### S5. Detailed results of the metalens with elliptical polarization focus.

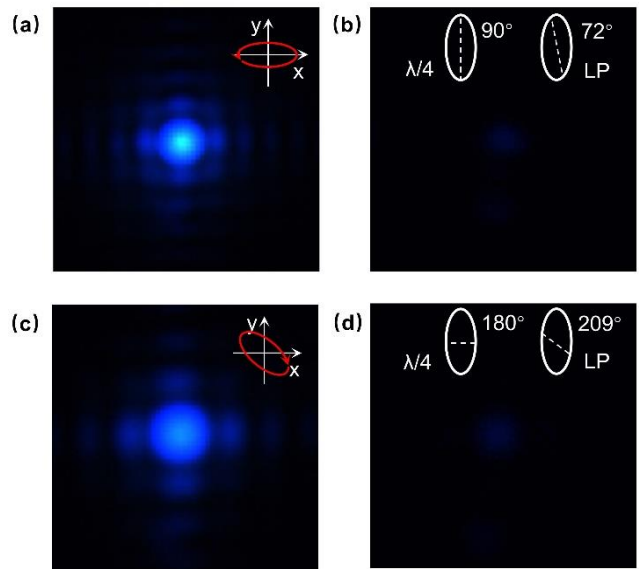

**Figure S5.** Polarization analyses of the focus with elliptical polarization. (a) Measured focal spot with polarization E on the Poincaré sphere. (b) Analyzed intensity profile with 90° QWP and 72° LP (0° corresponds to the transmitted axis in the x direction). (c) Measured focus with the polarization F on the Poincaré sphere. (d) Analyzed intensity profile with 180° QWP and 209° LP.

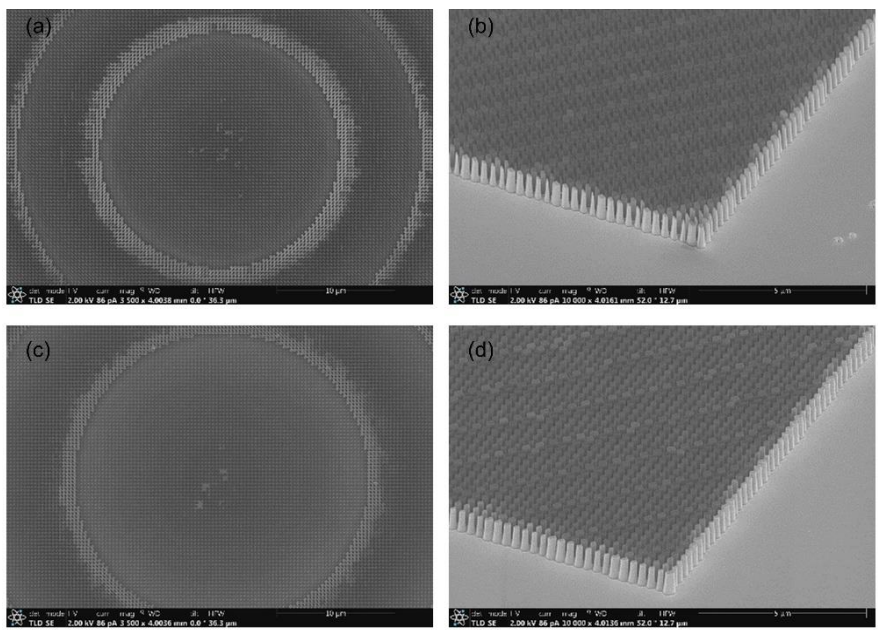

**Figure S6.** SEM images of the fabricated metalens with (a) the polarization E and (b) polarization F.